\documentclass[journal=jacsat,manuscript=article]{achemso}
\usepackage[version=3]{mhchem}
\usepackage{microtype}
\usepackage{graphicx}
\usepackage{caption}
\usepackage{hyperref}
\usepackage{amssymb}
\usepackage{lipsum} 

\usepackage{subcaption}
\usepackage{soul}
\hypersetup{
    colorlinks=true,
    linkcolor=blue,      
    citecolor=blue,       
    urlcolor=blue }    

\usepackage{graphicx}
\usepackage{amsmath}
\usepackage[table]{xcolor}
\sethlcolor{yellow} 
\usepackage{float}
\makeatletter 
\renewcommand*\acs@contact@details{%
  {\sffamily *Corresponding author: \acs@email@list}%
  \acs@number@list
}
\makeatother

\usepackage[most]{tcolorbox} 
\usepackage{booktabs}
\usepackage{makecell}
\usepackage{adjustbox}

\usepackage{natbib}

\title{Wafer-scale MoS$_\textbf{2}$ and HZO-based Ferroelectric Field Effect Transistors with 4.4 V Memory Window for Back-End Compatible Multibit Memory}

\author{Manisha Rajput}
\affiliation{Faculty of Information Technology and Communication Sciences, Tampere University, Tampere, Finland}

\author{Farhan Ahmed Siddique}
\affiliation{Faculty of Information Technology and Communication Sciences, Tampere University, Tampere, Finland}

\author{Chung-Che Huang}

\affiliation{Electronics and Computer Science, University of Southampton, Southampton, UK}
\author{Ioannis Zeimpekis}
\affiliation{Electronics and Computer Science, University of Southampton, Southampton, UK}
\author{Sayani Majumdar}
\affiliation{Faculty of Information Technology and Communication Sciences, Tampere University, Tampere, Finland}

\email{sayani.majumdar@tuni.fi}
\abbreviations{2D, CMOS, FeFET}

\begin{document}
\maketitle



\begin{abstract} 
 Ferroelectric field-effect transistors (FeFETs) with low-power operation and non-volatile data retention enabled by stable ferroelectric polarization switching, have emerged as promising alternatives to conventional floating-gate transistors for next-generation low-power memory applications. A FeFET with two-dimensional (2D) channel provides the opportunity for better electrostatic gate control and scalability beyond silicon.  In this work, we report wafer scale 2D semiconductor-based FeFETs  with Hf\textsubscript{0.5}Zr\textsubscript{0.5}O\textsubscript{2} (HZO) gate stack showing unusual p-type charge transport and stable multibit memory properties. FeFETs based on 2-3 monolayer Molybdenum disulfide (MoS\textsubscript{2}) channel and a 20 nm HZO back gate oxide layer exhibit a large hysteresis with a memory window spanning 4.4 V  under a operating voltage of $\pm$3.8 V, and an I\textsubscript{ON}/I\textsubscript{OFF} ratio of ~10\textsuperscript{5}. We demonstrate that the polarization charge-induced modification of Schottky barrier width strongly affects the hole transport and memory properties of MoS\textsubscript{2}-HZO FeFETs. We further demonstrate  several advantages and unique features of the FeFETs such as nondestructive data readout, non-volatile data retention under zero gate bias, low operation voltage, high write endurance and ultra-fast switching speed. These 1T memory cells can simplify the dense 3D integrated array architecture and offer additional practical functionalities, such as fast programming and erasing of large data arrays that are absent for many conventional memories. The demonstrated MoS\textsubscript{2} and HZO-based FeFET shows great potential for on-chip complementary metal-oxide semiconductor (CMOS)-back-end-of-line (BEOL) compatible circuits for multifunctional devices, especially for 3D monolithically integrated architectures for AI workload.

\end{abstract} 

\section{Introduction}

2D semiconductors have emerged as one most versatile class of beyond Silicon semiconductors due to their attractive electronic and optical properties.\cite{chhowalla2016two,10.1021/acsnano.3c12938,kim2024future,10.1021/acsnano.3c10900}  Key advantages of 2D semiconductor- based field-effect transistors (FETs) include superior gate control of channel conductance in scaled devices.\cite{chhowalla2016two,das2021transistors,zeng2024transistor} Atomic thickness of the channel layer gives the gate voltage a tight electrostatic control over the channel making them effective at small channel lengths. Also, high surface-to-volume ratio of 2D semiconductors reduce the off-state current leakage and pristine, dangling-bond-free surfaces oppose short-channel effects like drain-induced barrier lowering.\cite{WEI2022105160,pal2024three,sarkar2015subthermionic} From system level perspective, 2D devices can be integrated on any substrate, and this flexibility allows vertical monolithic 3D stacking at the BEOL with CMOS logic or sensing platforms. 

Molybdenum disulfide (MoS\textsubscript{2}) is one of the most studied 2D transition-metal dichalcogenides (TMDs), offering high electrical mobility at nanoscale thickness and good chemical stability.\cite{manzeli20172d,song2015synthesis} Monolayer MoS\textsubscript{2} is a direct-bandgap semiconductor with a bandgap of 1.9~eV, which transitions to an indirect-bandgap semiconductor with a bandgap of 1.3~eV as the thickness increases from monolayer to multilayer.\cite{PhysRevLett.105.136805} Recent studies have demonstrated the successful integration of 2D channel material-based FETs at the CMOS BEOL \cite{guan2023monolithic,wang2026two,kim2023scalable,10.1021/acsnano.6c08339} or on flexible substrates\cite{10.1021/acsnano.7b03703,li2020large,li2023large}, enabling the fabrication of complex logic circuits. However, stable non-volatile memory transistors based on 2D semiconductors remain relatively under-developed.
  
In recent years, ferroelectric memories emerged as a promising solution as embedded or standalone non-volatile memory (NVM) and for in-memory computing (IMC).\cite{yu2024ferroelectric,bhardwaj2026toward} Among the ferroelectric memories, ferroelectric random-access memories (FeRAM) is the most matured technology. Ferroelectric field-effect transistor (FeFET) memory also gained wide attention as a promising technology as it combines non-volatility, low-power operation, high-speed switching, and excellent scalability within a transistor-based architecture.\cite{kim2023wurtzite,khan2020future,wang2021van,yang2024emerging,li2026dual,10.1063/5.0278057} The polarization state of an ultrathin ferroelectric layer, can modulate the channel conductance, enabling voltage controlled device On and Off states. As polarization states remain stable even after removal of the applied bias, it enables information storage without continuous power consumption. In addition, FeFETs offer low write energy, nondestructive readout, and potentially high endurance, while their polarization-controllable intermediate states provide opportunities for multilevel storage and analog weight updates. These characteristics make FeFETs particularly attractive not only for conventional embedded non-volatile memory but also for IMC, neuromorphic hardware, and energy-efficient AI accelerators, where memory density, energy efficiency, and computation-memory integration are critical.\cite{10.1063/5.0035515,khan2020future,soliman2023first,8268338,10.1063/5.0278057} Despite these promises, Si-based FeFETs face several challenges that limit their practical scalability and performance reliability, especially under prolonged bias-stress.\cite{7599662,10557625,10.1021/acsami.6c05258,11353477,11066231,10.1063/5.0293794} A key issue is the integration of the ferroelectric layer with the Si channel, where interface dead layer formation between Si channel and ferroelectric layer leads to interface traps, defect states, and depolarization fields, degrading carrier mobility, threshold-voltage control, and memory window. 

Among the CMOS compatible ferroelectric materials, zirconium -doped hafnia oxide (Hf\textsubscript{0.5}Zr\textsubscript{0.5}O\textsubscript{2}, HZO) has emerged as a strong candidate.\cite{majumdar2022back} Its stable ferroelectric properties down to a few nanometer thickness, industry ready scalable process technology like atomic layer deposition (ALD) and low thermal budget fabrication made it an attractive choice  for memory applications. The compatibility of HfO\textsubscript{2}-based ferroelectrics with established CMOS processing technologies further enhances the potential for dense integration and advanced technology-node scaling. HZO’s orthorhombic (o-) phase with Pca21 structure has asymmetric charge center, that results in a net electric dipole moment inside the lattice, which under an applied electric field results in a switchable polarization, leading to a net polarization charge under applied electric fields. Ferroelectric HZO has been applied into novel low power consumption devices like negative capacitance transistors (NCFETs)\cite{jo2023negative,li2015sub,si2018steep,https://doi.org/10.1002/adfm.202003859}, ferroelectric tunnel junction (FTJs),\cite{10.1021/acsami.6b16173,guan2026probabilistic,10.1063/5.0029516,10.1021/acsami.1c18165} and FeFET memories.\cite{10.1063/1.5029324,li2024high,10.1021/acsnano.5b02078,li2026dual} However, in Si-based FeFETs, achieving a a suitable balance among different key performance indicators like large memory window, high endurance, long retention and multibit operation remains a major challenge, the high endurance being especially a critical issue. These limitations are particularly important for neuromorphic and IMC applications, where precise and reproducible conductance modulation and large write endurance are required for precise synaptic weight update rather than simply storing binary states.   
Integration of FE materials with low thermal budget semiconductors is not very straight forward either. For 2D semiconductors, challenge arise because of the high density of interface states at the FE/2D interface. These states can completely screen the ferroelectric polarization charge, thereby decoupling the ferroelectric from the 2D channel. Generally, trap-related hysteresis in the transfer characteristics occurs in the opposite direction to ferroelectric-induced hysteresis, reducing the FE-related hysteresis window or causing hysteresis in an undesirable direction. Consequently, trap-related hysteresis has frequently been reported in FeFETs,\cite{Sun_2017,8667000,lipatov2021using,10.1021/acsnano.5b02078} whereas only a limited number of studies have reported hysteresis arising purely from ferroelectric polarization switching.\cite{10.1021/acs.nanolett.9b00180,10.1021/acsnano.4c11081} Furthermore, some 2D ferroelectric transistors have shown no hysteresis, possibly because the ferroelectric film acts primarily as a high-\(\kappa\) dielectric or trap and polarization related hysteresis, that is opposite in direction, compensate each other.\cite{10.1039/c5nr01072a,pucher2026mechanisms} Therefore, FE/2D heterostructures in which the transistor characteristics are controlled primarily by the ferroelectric polarization charge are desirable. 

A 2D channel material in a FeFET can bring memory window stability and scalability by removing the critical interface dead layer problems that plague traditional silicon-based FeFETs.\cite{Huang_2024}It is envisioned that 2D channels can eliminate interfacial oxide layers as 2D materials have pristine, dangling-bond-free surfaces. This can prevent the formation of an unwanted amorphous oxide layer between the channel and the ferroelectric layer, which otherwise reduces the effective voltage drop across the ferroelectric film leading to maximized remnant polarization (Pr). The atomic thinness of the 2D channel ensures that the entire channel experiences a uniform, high electric field from the ferroelectric gate. Eliminating dangling bonds at the FE-semiconductor interface can minimize charge trapping at the interface, improving ferroelectric MW, suppressing leakage currents during write cycles and improving endurance. The ultra-thin nature of 2D materials reduces the internal depolarization field, which directly prevents state retention loss over time. Excellent electrostatic control combined with ideal ferroelectric coupling creates a highly pronounced difference between the Off and On states. Low-temperature processing allows these devices to be vertically integrated in BEOL configurations for high-density storage. However, in reality, several practical implementation issues exist. Achieving a large and stable memory window in a 2D FeFET remained a challenge. So far a few works reported MoS\textsubscript{2} based FeFETs with memory windows spanning from ~0.1–0.22 V in low-voltage devices\cite{10413873} to ~3.6 V\cite{10.1021/acsnano.6c08339} in interface-engineered monolayer-MoS\textsubscript{2} FeFETs.\cite{zhang2020low} While the small voltage operating devices are suitable for advanced node logic integration with binary operation, achieving analog or multibit states would be challenging in these devices. To achieve large MW FeFETs \citeauthor{10.1021/acsnano.6c08339} introduces, a Y\textsubscript{2}O\textsubscript{3} dielectric layer in the gate stack which as a consequence degraded the retention by increasing the depolarization field. Our work here, report on a multibit FeFET based on HZO gate oxide demonstrating p-type MoS\textsubscript{2} channel with a maximum memory window of 4.4 V. These devices present high stability over multiple bits without any signs of decay over a very long time, endurance over 10\textsuperscript{7} switching cycles and nanosecond range of programming and erasing time. In all previous reports, a monolayer MoS\textsubscript{2} is used as the channel with devices showing n-type conduction and hysteresis. In those FeFETs, charge trapping at the dielectric-semiconductor interface resulted in degraded memory performance.  Here, we show that pairing a suitable 2D channel with a ferroelectric gate stack enables high-performance p-type FeFETs, with an on/off current ratio of $10^5$, a memory window of 4.4 V at a operating voltage of $\pm$3.8 V, endurance upto $10^7$ switching cycles, 16 programmable states with stable retention, and fast programming with pulses as short as 100 ns. Notably we show stable p-type FeFET operation from wafer-scale MoS\textsubscript{2} based devices for the first time where both MoS\textsubscript{2} and HZO are fabricated with industry-compatible ALD processes. These wafer-scale FeFETs shows highly stable multibit NVM performance, suitable for non-volatile data storage  and in IMC architectures whilst significantly opening the prospect of complementary functionality lowering energy requirements.

\section{Results and discussion}\label{sec2}

\subsection{Device Architecture and Structural Characterization of MoS$_2$}
\label{results} 
\begin{figure}
    \centering
\includegraphics[width=1\textwidth]{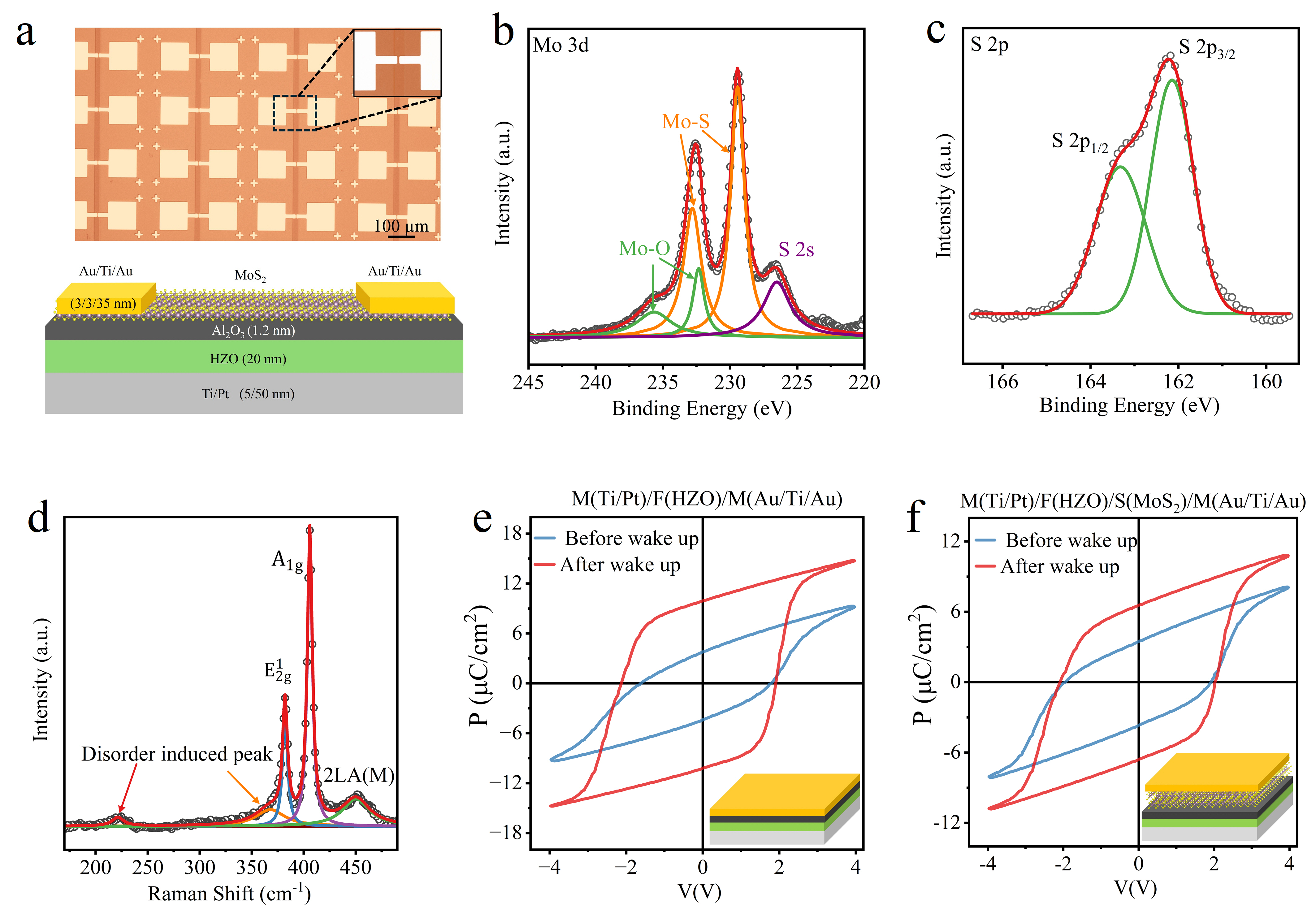}
     \caption{(a) Schematic of the 2D FeFET incorporating three-layer MoS\textsubscript{2} as the semiconducting channel and 20~nm HZO as a back-gate dielectric. (b) XPS of Mo 3d core spectra. (c) XPS of S 2s core spectra. (d) Raman spectra of MoS\textsubscript{2}. (e) Polarization--voltage (P--V) characteristic of HZO in the MFM structure before and after wake-up. (f) Polarization--voltage (P--V) characteristic of HZO in the MFS structure before and after wake-up.}
    \label{fig:1}
\end{figure}

\textbf{Figure}\ref{fig:1}a schematically illustrates the device architecture of the proposed p-FeFET, consisting of a 20~nm HZO bottom-gate dielectric deposited by atomic layer deposition (ALD) on a Ti/Pt gate electrode. Following HZO deposition, an Al\textsubscript{2}O\textsubscript{3} capping layer (1~nm ) was deposited at \(200^\circ\)C without breaking vacuum. For fabrication of the MoS\textsubscript{2} FeFETs, the ALD-grown trilayer MoS\textsubscript{2} film was transferred from the Si/SiO\textsubscript{2} (300 nm) growth substrate onto the 20~nm HZO dielectric using a polymethyl methacrylate (PMMA)-assisted wet-transfer process. The MoS\textsubscript{2} channel regions were then defined by photolithography followed by reactive ion etching (RIE) using SF\textsubscript{6} chemistry. Finally, source and drain contacts were patterned by photolithography, followed by evaporation of a 3/3/35 nm Au/Ti/Au metal stack and a subsequent lift-off process. To improve the metal-semiconductor contact properties vacuum thermal annealing at 300 $^\circ$ C for 30 minutes was performed. Detailed information on the MoS\textsubscript{2} transfer and device fabrication is provided in the Methods section and Supplementary Figure S1. Figures \ref{fig:1}b and \ref{fig:1}c present the Mo 3d and S 2p core-level spectra of the MoS\textsubscript{2} film, respectively. The Mo 3d spectrum exhibits the Mo(IV) 3d\textsubscript{3/2} and 3d\textsubscript{5/2} peaks at 233.7 and 229.4 eV, respectively, along with the S 2s peak at 226.5 eV. In addition, two well-pronounced peaks at 235.6 and 232.3 eV are observed in the Mo 3d spectrum, which can be attributed to Mo(VI) 3d\textsubscript{3/2} and 3d\textsubscript{5/2}, respectively. These Mo(VI) components indicate the presence of MoO\textsubscript{x} species and suggest the formation of Mo--O bonds in the MoS\textsubscript{2} film. The observed Mo--O bonds may originate from oxygen-passivated sulfur vacancies in MoS\textsubscript{2} or residual MoO\textsubscript{x} clusters formed during growth. The presence of MoO\textsubscript{x} species has been reported to induce p-type doping in MoS\textsubscript{2}.\cite{10.1063/1.4983092} Previous studies have shown that the formation of Mo--O bonds at sulfur vacancy sites heals the donor states and introduces acceptor states near the valence-band edge, leading to p-type behaviour.\cite{Shen2022Healing,10.1021/nn500532f,10.1063/1.4983092} Similarly, oxygen plasma treatment has been used to convert the to convert top few layers of TMDs into native substoichiometric oxide layers, resulting in p-type doping through surface charge transfer.\cite{10.1021/acsnano.0c05572}
Figure \ref{fig:1}d shows the Raman spectrum of the MoS\textsubscript{2} film, showing the in-plane E\textsuperscript{1}\textsubscript{2g} vibration mode at \(\sim382\) cm\(^{-1}\) and the out-of-plane A\textsubscript{1g} vibration mode at \(\sim405.8\) cm\(^{-1}\). The additional peaks observed at \(\sim221\) cm\(^{-1}\) and \(\sim360\) cm\(^{-1}\), together with the longitudinal acoustic mode (2LA(M)) at \(\sim450\), are associated with disorder in the MoS\textsubscript{2} film.\cite{PhysRevB.91.195411} The two main components of the 2D FeFET are the ferroelectric gate stack and the 2D channel, as illustrated in Figure \ref{fig:1}a. Before integrating the ferroelectric gate stack with the 2D FET, it is important to understand the polarization switching characteristics of the gate stack. To investigate the polarization switching behavior of the 20 nm HZO gate stack, metal-ferroelectric-metal (MFM) and metal-ferroelectric-semiconductor-metal (MFSM) capacitor structures were fabricated, as shown in the insets of Figures \ref{fig:1}e and \ref{fig:1}f, respectively. For the MFM structure, the bottom Ti/Pt (5/50~nm) electrode was deposited by e-beam evaporation, followed by the deposition of HZO (20~nm) and Al\textsubscript{2}O\textsubscript{3} (1.2~nm) layers using ALD. The top Au/Ti/Au (3/3/35~nm) metal stack was subsequently deposited by e-beam evaporation. For the MFSM structure, the ALD-grown trilayer MoS\textsubscript{2} film was transferred onto the HZO/Al\textsubscript{2}O\textsubscript{3} stack using a PMMA-assisted wet-transfer process, followed by the deposition of the Au/Ti/Au (3/3/35~nm) metal stack. To induce the ferroelectric phase in the HZO layer, thermal annealing was performed at 350 $^\circ$ for 30 min in a nitrogen (N$_2$) environment. This step is crucial for enhancing the ferroelectric properties of the HZO Figures \ref{fig:1}e and \ref{fig:1}f show the typical polarization-voltage (P-V) characteristics of the MFM and MFSM structures, respectively, before and after the wake-up process. Following wake-up, an increase in the polarization values are observed for both structures. For the MFSM structure, remanent polarization (P\textsubscript{r}) value decreases from approximately 10 to 8~$\mu$C/cm$^{2}$ compared with the MFM structure. This reduction in polarization may be associated with the integration of the MoS\textsubscript{2} semiconductor layer, in absence of a gate bias is insulating and can induce depolarizing field  which can influencing the local electric field environment and polarization charge screening.

\vspace{10pt}

\subsection{Electrical Performance of HZO-gated MoS$_2$ FeFETs}

\begin{figure}[H]
    \centering
    \includegraphics[width=1\linewidth]{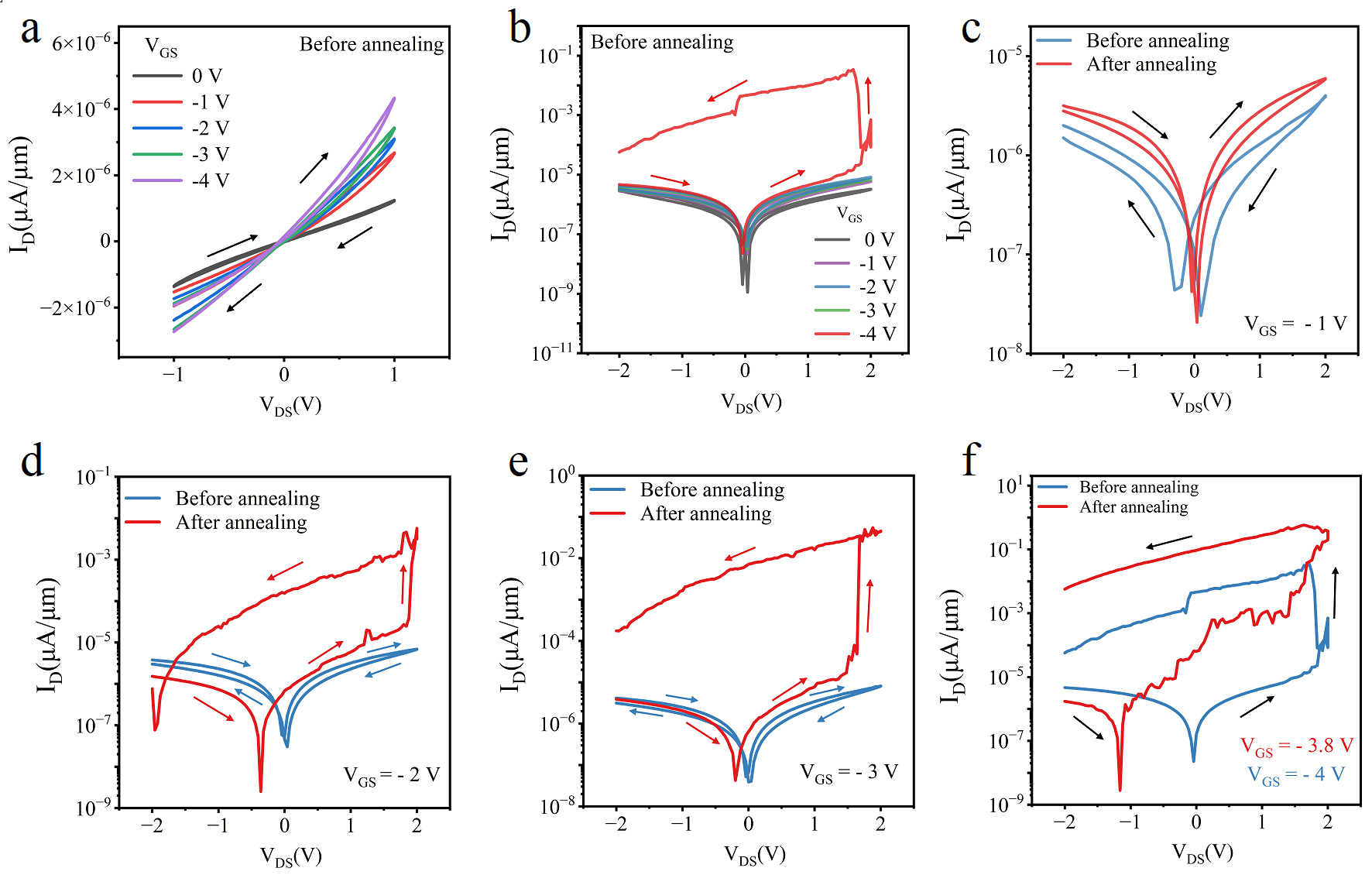}
    \caption{Output characteristics of the MoS\textsubscript{2} p-FeFET before and after thermal annealing of the ferroelectric gate dielectric. (a) $I_{\mathrm{D}}$-$V_{\mathrm{DS}}$ characteristics before annealing at different $V_{\mathrm{GS}}$ obtained by dual sweeping $V_{\mathrm{DS}}$ between $\pm1~\mathrm{V}$. (b) Semi-log plot of the $I_{\mathrm{D}}$-$V_{\mathrm{DS}}$ characteristics before annealing at different $V_{\mathrm{GS}}$, obtained by dual sweeping $V_{\mathrm{DS}}$ between $\pm2~\mathrm{V}$. (c-f) Comparison of the $I_{\mathrm{D}}$-$V_{\mathrm{DS}}$ characteristics plotted on a semi-logarithmic scale before and after thermal annealing at $V_{\mathrm{GS}}=-1$, $-2$, and $-3~\mathrm{V}$ for (c-e), respectively, and at $V_{\mathrm{GS}}=-4~\mathrm{V}$ before thermal annealing and $-3.8~\mathrm{V}$ after thermal annealing for (f),the blue and red curves represent the data before and after RTA, respectively.}
    \label{fig:2}
\end{figure}

Ferroelectric polarization generates an utltra-high electric field leading to substantial modulation of carrier concentration and band structure within the channel.\cite{Wang2015MoS2Ferroelectric,sun2024ambipolar} To examine the effect of polarization we investigated FET characteristics before and after performing thermal annealing of gate dielectric. Figure \ref{fig:2}a shows the output characteristics i.e., the source-to-drain current  (I\textsubscript{D}) as a function of source-to-drain voltage (V\textsubscript{DS}) for various back gate voltages (V\textsubscript{GS}) before annealing. As can be seen, the I\textsubscript{D}--V\textsubscript{DS} curves deviate from linearity, suggesting the presence of a Schottky barrier at the metal/MoS\textsubscript{2} interface. Furthermore, Figure~\ref{fig:2}b presents the output characteristics obtained by dual-sweeping $V_{\textsubscript{DS}}$ between $\pm 2~\mathrm{V}$. Notably, at $V_{\textsubscript{GS}}=-4~\mathrm{V}$, $I_{\textsubscript{D}}$ increases by more than three orders of magnitude, accompanied by a large hysteresis and a reversal of the hysteresis direction from clockwise to counter-clockwise. Similarly, we studied the $I_{\textsubscript{D}}$-$V_{\textsubscript{DS}}$ characteristics after thermal annealing of the gate dielectric. Figure~\ref{fig:2}c-f shows a comparison of the $I_{\textsubscript{D}}$-$V_{\textsubscript{DS}}$ characteristics before and after annealing. Following annealing, a sharp rise in the drain current and a transition in the hysteresis direction are observed at $V_{\textsubscript{GS}}=-2~\mathrm{V}$ (Fig.~\ref{fig:2}d), compared with $V_{\textsubscript{GS}}=-4~\mathrm{V}$ before annealing (Fig.~\ref{fig:2}b). At each $V_{\textsubscript{GS}}$, the drain current is significantly higher after annealing, accompanied by a wider hysteresis window.

\begin{figure}[H]
    \centering
    \includegraphics[width=1\linewidth]{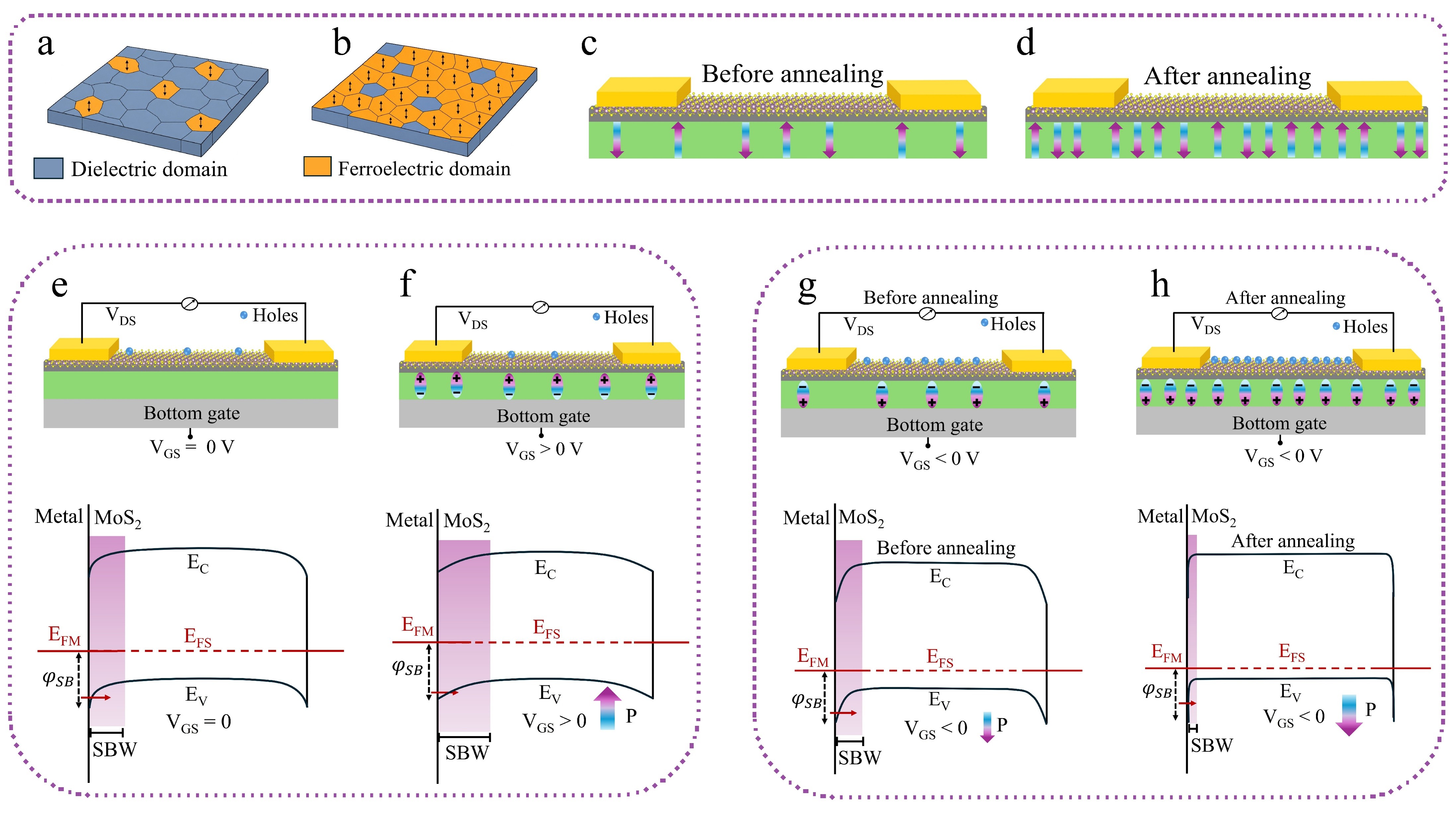}
    \caption{Polarization-induced modulation of the Schottky barrier width and carrier density. (a,b) Schematics of the ferroelectric dielectric (a) before and (b) after annealing, showing an increased fraction of ferroelectric domains after annealing due to crystallization. (c,d) Corresponding device schematics (c) before and (d) after annealing, illustrating the lower and higher net polarization, respectively. (e) Schematic of the MoS$_2$ p-FeFET at $V_{\mathrm{GS}}=0$, showing a relatively low hole concentration in the channel, together with the corresponding energy band diagram showing a wider Schottky barrier at the metal--MoS$_2$ contact. (f) Schematic of the device under positive gate bias, showing hole depletion in the channel due to upward polarization, together with the corresponding energy band diagram illustrating a wider Schottky barrier. (g) Schematic of the device before annealing under negative gate bias, showing hole accumulation in the channel due to downward polarization, together with the corresponding energy band diagram illustrating a reduced Schottky barrier width. (h) Schematic of the device after annealing under negative gate bias, showing greater hole accumulation due to the larger net downward polarization, together with the corresponding energy band diagram illustrating a further reduction in the Schottky barrier width.}
    \label{fig:3}
\end{figure}

The distinct output characteristics observed before and after thermal annealing can be attributed to changes in the ferroelectric polarization of the HZO layer. Before thermal annealing, the HZO layer contains a smaller fraction of ferroelectric domains (Fig.~\ref{fig:3}a,c) due to partial formation of the orthorhombic phase during vacuum annealing at 300~$^\circ$C for 30~min, performed to improve the metal--semiconductor contact resistance. Thermal annealing increases the fraction of the ferroelectric orthorhombic phase, leading to a larger population of ferroelectric domains (Fig.~\ref{fig:3}b) and consequently a higher net polarization, as represented in Fig.~\ref{fig:3}d. 
The Schottky barrier width at the metal--semiconductor contact can be estimated using the depletion approximation:

\begin{equation}
W = \sqrt{\frac{2\varepsilon_s(V_{\mathrm{bi}}-V)}{qN_A}},
\end{equation}

where $\varepsilon_s$ is the semiconductor permittivity, $V_{\mathrm{bi}}$ is the built-in potential, $V$ is the applied bias, $q$ is the elementary charge, and $N_A$ is the local doping concentration. This expression indicates that a lower hole concentration leads to a wider depletion region, whereas a higher hole concentration narrows the depletion region and reduces the effective Schottky barrier width. Indeed, doping the semiconductor near the metal contact has been reported as an effective approach to mitigate Schottky barrier effects and reduce contact resistance~\cite{10.1021/nl502603d,10.1021/nl301702r,10.1021/acsnano.0c05572,10.1021/acs.nanolett.5b05066}. The increased carrier concentration at the contact region due to ferroelectric polarization induced charges in our p-FeFETs narrows the effective Schottky barrier width, thereby increasing the probability of carrier tunnelling across the metal--semiconductor (M--S) interface leading to increase in drain current near the ferroelectric coercive voltage and non-volatile change in drain current due to polarization remanence. To understand the distinct output characteristics observed under different gate-bias conditions before and after thermal annealing (Fig.~\ref{fig:2}), Fig.~\ref{fig:3}e--h schematically illustrates how ferroelectric polarization modulates the carrier density in the MoS$_2$ channel and the Schottky barrier width at the metal--semiconductor contact. At zero gate bias, the weakly p-type MoS$_2$ channel has a relatively low hole concentration (Fig.~\ref{fig:3}e). The low hole concentration near the metal--semiconductor contact results in a wider depletion region and a wider Schottky barrier, suppressing tunnelling-assisted hole injection. Under positive gate bias, the spontaneous polarization of HZO points towards ($P_{\uparrow}$) the MoS$_2$ channel, inducing further hole depletion to screen the polarization field (Fig.~\ref{fig:3}f). The resulting increase in Schottky barrier width further suppresses hole injection, leading to a lower drain current than at $V{\mathrm{GS}}=0$ V, both before and after thermal annealing (Supplementary Information, Figure S3).  Under negative gate bias, the HZO polarization points downward ($P_{\downarrow}$), generating negative bound charge at the HZO/MoS$2$ interface and leading to hole accumulation in the MoS$_2$ channel (Fig.~\ref{fig:3}g,h). The increased hole concentration near the contact thins the Schottky barrier width, facilitating tunnelling-assisted hole injection and increasing the drain current. Before RTA, the lower net polarization of HZO induces weaker hole accumulation, requiring a more negative $V_{\mathrm{GS}}$ to narrow the Schottky barrier width sufficiently for efficient hole injection (Fig.~\ref{fig:3}g). Accordingly, the sharp rise in drain current and the reversal of the hysteresis direction occur at $V_{\mathrm{GS}} = -4$ V (Fig.~\ref{fig:2}b).

\begin{figure}[H]
    \centering
    \includegraphics[width=1\linewidth]{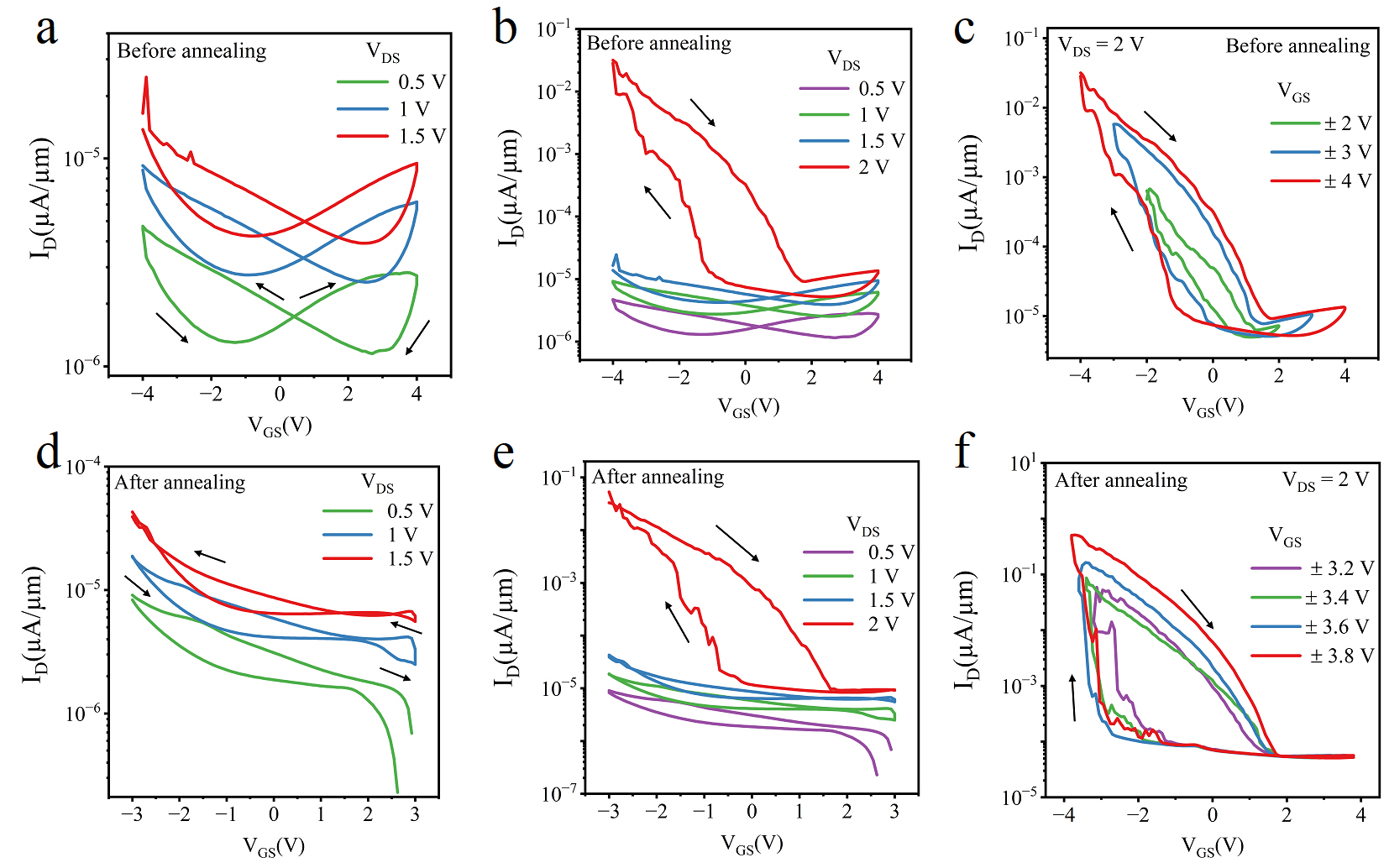}
    \caption{Electrical characterization of the MoS\textsubscript{2} FeFET before and after annealing of the gate dielectric. (a) Transfer characteristics before annealing measured by sweeping $V_{\mathrm{GS}}$ over a range of $\pm4$~V at $V_{\mathrm{DS}}=0.5$--$1.5$~V in steps of 0.5~V. (b) Transfer characteristics before annealing measured by sweeping $V_{\mathrm{GS}}$ from $\pm4$~V at $V_{\mathrm{DS}}=0.5$--$2$~V in steps of 0.5~V. (c) Transfer characteristics before annealing measured by sweeping $V_{\mathrm{GS}}$ over $\pm2$, $\pm3$, and $\pm4$~V at $V_{\mathrm{DS}}=2$~V. (d) Transfer characteristics after annealing measured by sweeping $V_{\mathrm{GS}}$ from $\pm3$~V at $V_{\mathrm{DS}}=0.5$--$1.5$~V in steps of 0.5~V. (e) Transfer characteristics after annealing measured by sweeping $V_{\mathrm{GS}}$ from $\pm3$~V at $V_{\mathrm{DS}}=0.5$--$2$~V in steps of 0.5~V. (f) Transfer characteristics after annealing measured by sweeping $V_{\mathrm{GS}}$ over $\pm3.2$, $\pm3.4$, $\pm3.6$, and $\pm3.8$~V at $V_{\mathrm{DS}}=2$~V.}
    \label{fig:4}
\end{figure}

After thermal annealing the larger net ferroelectric polarization promotes hole accumulation at less negative gate voltages, narrowing the Schottky barrier width (Fig.~\ref{fig:3}h), enabling more hole injection at lower $|V_{\mathrm{GS}}|$. Consequently, the sharp rise in drain current and the reversal of the hysteresis direction occur at  $V_{\mathrm{GS}} = -2$ V (Fig.~\ref{fig:2}d) with a significantly higher drain current across the measured $V_{\mathrm{GS}}$ range. Next, we investigated the transfer characteristics before and after RTA. Figures~\ref{fig:4}a-b show the transfer characteristics obtained by dual sweeping $V_{\mathrm{GS}}$ between $\pm4$ V before annealing. At $V_{\mathrm{DS}}=0.5$--$1.5$~V, both p-type and n-type branches are observed, with anticlockwise and clockwise hysteresis, respectively. The observed hysteresis directions are indicative of trap-related effects. The $I_{\mathrm{ON}}/I_{\mathrm{OFF}}$ ratio remains below one order of magnitude for both branches, indicating weak electrostatic modulation.  At $V_{\mathrm{DS}}=2$~V, the transfer characteristics show pronounced p-type transport with clockwise hysteresis and an $I_{\mathrm{ON}}/I_{\mathrm{OFF}}$ ratio of approximately $10^4$ with the memory window of 1.8 V. We attribute this evolution of the transfer characteristics at $V_{\mathrm{DS}}=2$~V to injected holes which facilitate the screening of the interfacial bound charge. At lower $V_{\mathrm{DS}}$, the hole injection is insufficient to effectively screen the polarization-induced bound charge at the MoS$_2$/HZO interface. At $V_{\mathrm{DS}}=2$~V, the injected holes provide effective screening of the interfacial bound charge, thereby reducing the depolarization field and stabilizing the downward polarization state in HZO. The resulting polarization-induced electric field promotes substantial hole accumulation in the MoS$_2$ channel, which narrows the Schottky barrier width and facilitates hole injection, as discussed above. Consequently, hole conduction becomes dominant over electron conduction, resulting in the transition from weak ambipolar to predominantly p-type transport at $V_{\mathrm{DS}}=2$~V. The change in the p-type branch hysteresis direction from anticlockwise to clockwise at \(V_{\mathrm{DS}} = 2\) V indicates  shift from trap-related to ferroelectric-dominated hysteresis.

  Additionally, the transfer characteristics measured at different scan rates and under different gate-bias stress durations are presented in Supplementary Figure S4. No significant changes are observed under either condition, indicating a minor contribution from charge trapping. Furthermore, Figure \ref{fig:4}c illustrates the transfer characteristics by  dual sweeping the $V_{\mathrm{GS}}$ for different ranges $V_{\mathrm{DS}}=2$~V  before annealing. Predominantly p-type transport with clockwise hysteresis is observed for the narrower $V_{\mathrm{GS}}$ sweep ranges, further supporting the proposed role of ferroelectric polarization at $V_{\mathrm{DS}}=2$~V in modulating the transport characteristics.  Figures~\ref{fig:4}d-f present the transfer characteristics obtained by dual sweeping $V_{\mathrm{GS}}$ between $\pm3$ V after thermal annealing of the gate dielectric. At $V_{\mathrm{DS}}=0.5$--$1.5$~V, a weak p-branch with anticlockwise hysteresis is observed, while the n-branch is completely suppressed. At $V_{\mathrm{DS}}=2$ V, the transport becomes p-type, similar to that observed before annealing, with a one-order-of-magnitude increase in the hole current. Upon increasing the $V_{\mathrm{GS}}$ sweep range, the hysteresis becomes more pronounced, resulting in an enhanced memory window (Figure~\ref{fig:4}f). The memory window reaches 4 V at a $V_{\mathrm{GS}}$ sweep range of $\pm3.8$ V, whereas it was 1.8 V at a $V_{\mathrm{GS}}$ sweep range of $\pm4$ V before annealing (Figure~\ref{fig:4}c). 
 
\begin{figure}[H]
    \centering
    \includegraphics[width=1\linewidth]{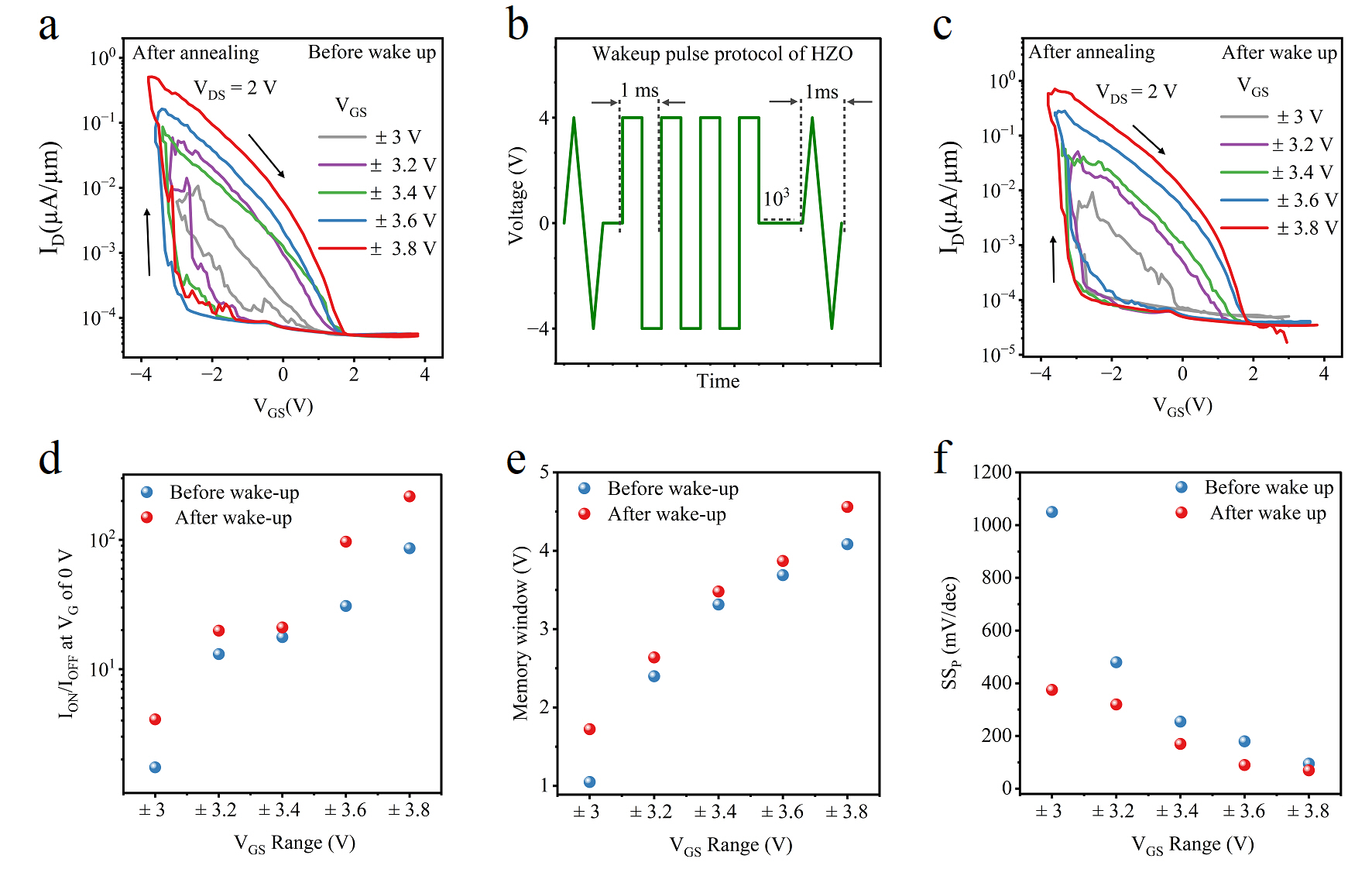}
    \caption{Impact of the wake-up of gate dielectric on device metrics (a) Transfer characteristics measured at different $V_{\mathrm{GS}}$ sweep ranges from  V before wake-up. (b) Wake-up pulse protocol applied to the HZO gate dielectric. (c) Transfer characteristics measured after the wake-up process. (d) $I_{\mathrm{ON}}/I_{\mathrm{OFF}}$ ratio at $V_{\mathrm{GS}}=0$ V for different $V_{\mathrm{GS}}$ sweep ranges before and after wake-up. (e) Memory window extracted from the transfer characteristics before and after wake-up. (f) Subthreshold swing extracted over one order of magnitude change in $I_{\mathrm{D}}$ before and after wake-up.
}
    \label{fig:5}
\end{figure}

 MoS\textsubscript{2}-based FETs typically exhibit strong n-type behaviour, commonly attributed to donor states associated with sulfur vacancies and Fermi-level pinning (FLP) near the conduction band edge.\cite{10.1039/c7cs00828g} This favours electron injection, enabling high-performance n-type FETs using low work-function metals such as Sc, Bi and Sb, which provide near-zero Schottky barrier heights for electrons.\cite{shen2021ultralow,li2023approaching,das2013high} Conversely, the large Schottky barrier for hole injection makes p-type transport difficult to achieve in MoS\textsubscript{2}. High-work-function metals such as Pt and Pd have been explored as p-type contacts to MoS\textsubscript{2}, but persistent FLP still limits hole injection. Additionally, the high melting points of these metals can lead to damage during the deposition process. Several approaches have been explored to achieve p-type transport in MoS\textsubscript{2}, including oxygen plasma doping, electric double layer transistors (EDLT)\cite{10.1021/nl400902v}, contact engineering\cite{10.1021/nl4043505,10.1063/1.5100154}, charge transfer doping \cite{min2018charge}, substitutional doping \cite{10.1063/1.4919565} as well as substrate and dielectric effect. Achieving comparable performance in p-type MoS$_2$ FETs, however, demands more than the  channel and contact engineering. FeFETs offer an additional means of controlling carrier transport through polarization-induced modulation of the electrostatic potential at the semiconductor channel and its contacts. The p-type behaviour observed in our MoS\textsubscript{2}/HZO FeFETs can be attributed to two complementary mechanisms. First, oxygen incorporation in MoS\textsubscript{2} passivates donor states associated with sulfur vacancies and introduce additional acceptor-like states near the valence-band edge.\cite{Shen2022Healing} Although these can reduce the n-type doping they do not necessarily result in p-type devices. Our oxygen modification enables hole transport in our FeFET devices. Second, ferroelectric polarization in HZO strongly modulates the carrier concentration in the MoS\textsubscript{2} channel and the Schottky barrier width at the metal/MoS\textsubscript{2} contact, further enhancing p-type behavior. Next, we assessed the impact of the of larger polarization charge from gate dielectric with the wake-up protocol on the transfer characteristics. The transfer characteristics measured over different $V_{\mathrm{GS}}$ sweep ranges before and after wake-up are shown in Figures~\ref{fig:5}a and c, respectively, while Figure~\ref{fig:5}b shows the wake-up pulse protocol applied to the HZO.

 After wake-up, the hysteresis becomes more pronounced, particularly at lower $V_{\mathrm{GS}}$ sweep ranges. The resulting changes in the key metrics are summarized in Figures~\ref{fig:5}d-f. The $I_{\mathrm{ON}}/I_{\mathrm{OFF}}$ ratio at $V_{\mathrm{GS}}=0$ V and the memory window both increase after wake-up, as shown in Figures~\ref{fig:5}d and e, respectively. After wake up subthreshold swing (SS\textsubscript{P}), average over one order of magnitude change in $I_{\mathrm{D}}$, is improved reaching to a minimum value of 76~mV/dec (Figure~\ref{fig:5}f). Notably, all these improvements are more visible for the lower sweep range ($\pm3$ V), demonstrating its potential for low-voltage operation. Further reduction in the operating voltage could be achieved by increasing the electric field across gate dielectric. The influence of scan rate and gate-bias stress duration on the transfer characteristics, examined before and after wake-up, is presented in Supplementary Figures S5, respectively. In both cases, the transfer characteristics exhibit negligible variation.

\subsection{Statistical Analysis and Benchmarking of p-Type MoS$_2$ FETs} 

\begin{figure}[H]
    \centering
    \includegraphics[width=1\linewidth]{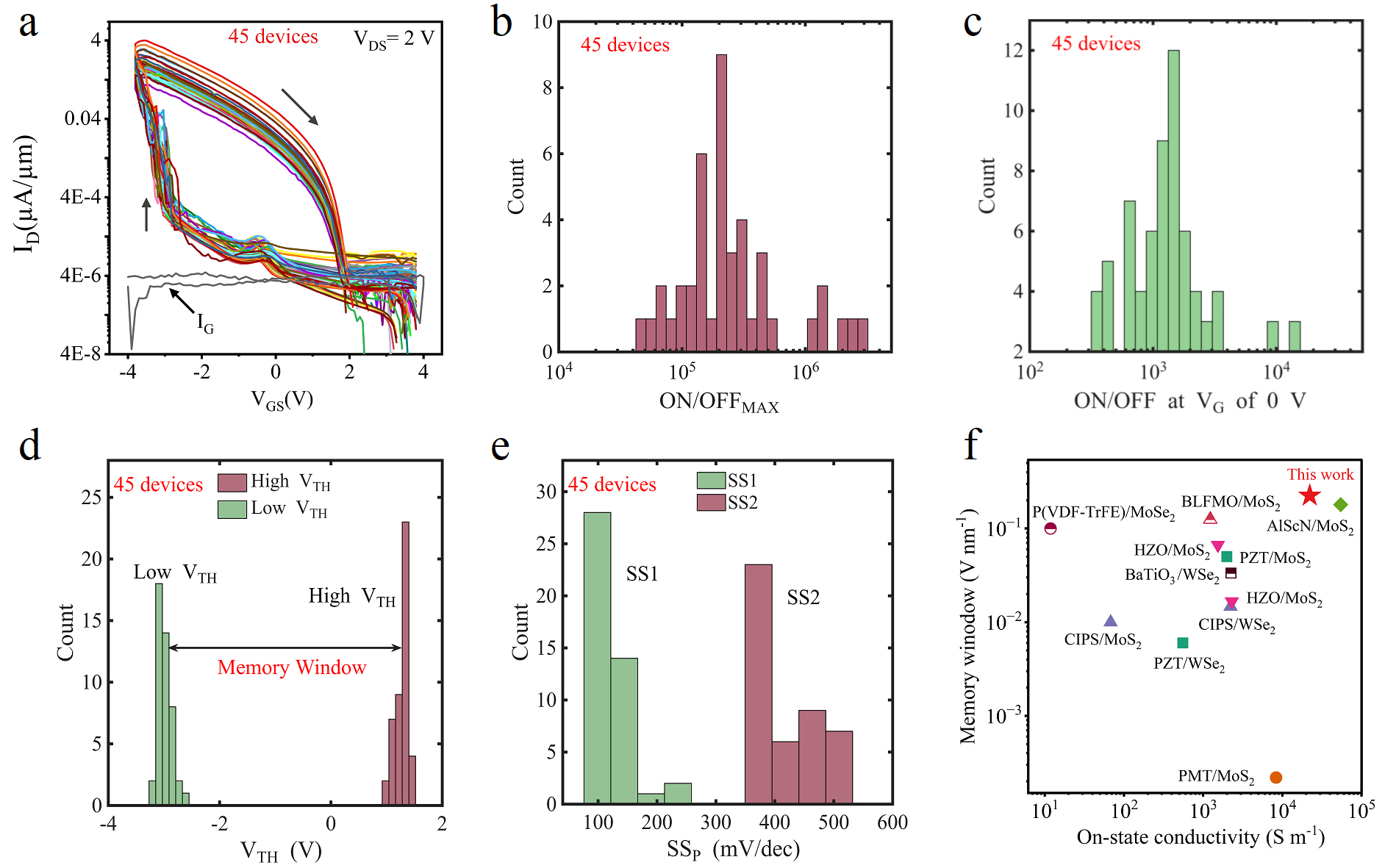}
    \caption{Statistical analysis and benchmarking of MoS$_2$ p-FeFETs.(a)Transfer characteristics of 45 devices with a channel length ($L$) of 5~$\mu$m and channel width ($W$) of 25~$\mu$m, measured at $V_{\mathrm{DS}}=2$~V with a $V_{\mathrm{GS}}$ sweep range of $\pm3.8$~V.(b) Distribution of the maximum ON/OFF ratio (ON/OFF$_{\mathrm{MAX}}$) extracted from the transfer characteristics, with a median value of $5\times10^{5}$. (c) Distribution of the ON/OFF ratio at $V_{\mathrm{GS}}=0$~V. (d) Distributions of the LVT and  and HVT extracted at $I_{\mathrm{D}}=10^{-8}$~A. (e) Distribution of the  $SS_{\mathrm{P}}$ extracted from the two sweep directions, averaged over two orders of magnitude change in ID, denoted as $SS_1$ and $SS_2$. (f) Comparison of the normalized MW and normalized ON-state conductivity of the MoS$_2$ p-FeFETs with previously reported 2D-channel FeFETs \cite{10.1021/acsnano.5b02078, ko2016ferroelectrically,10.1021/acsami.0c09951,10.1021/acsaelm.1c00492,10.1021/acsnano.8b01810,9178419,zhang2020low,kim2023scalable,10.1021/acs.nanolett.4c02069,wang2017ferroelectric,zhang2022tunable,10.1021/acsnano.5b04592}.
 }
    \label{fig:6} 
\end{figure}

 Further, the device metrics were evaluated across 45 devices to assess the reproducibility. The devices show significant overlap in their transfer characteristics (Figure~\ref{fig:6}a), for large-scale array integration. The FeFETs exhibit a ON/OFF\textsubscript{MAX} ratio exceeding $10^{5}$ (Figure~\ref{fig:6}b), while the ON/OFF ratio at $V_{\mathrm{GS}}=0$~V remains above $10^{3}$ (Figure~\ref{fig:6}c). Figure~\ref{fig:6}d summarizes the distributions of the low-threshold-voltage (LVT) and high-threshold-voltage (HVT) states, extracted at a drain current of $10^{-8}$~A. MW is calculated as the difference between the HVT and LVT values extracted at the same current level. The devices exhibit a median MW of approximately 4.4~V (Figure~\ref{fig:6}d). Figure~\ref{fig:6}e summarizes the subthreshold swing (SS\textsubscript{P}) extracted from the two sweep directions, denoted as $SS_1$ and $SS_2$, with a mean value of 108 and 425 mV/dec. Small variations in the device metrics of the MoS$_2$ FETs can be observed across different devices. These variations may arise from residues and wrinkles introduced during fabrication and transfer processes, as well as channel inhomogeneities. Figure~\ref{fig:6}f compares the normalized MW and ON-state conductivity of our devices with those extracted from previously reported 2D-channel FeFETs. Considering the dependence of the MW on the ferroelectric thickness, we use the thickness-normalized MW as a comparative metric. The ON-state conductivity is similarly normalized to the channel width/length and drain voltage. As evident from Figure~\ref{fig:6}f, our MoS$_2$ p-FeFETs demonstrate among the highest MW and ON-state conductivity values compared with previously reported 2D-channel FeFETs.

\begin{figure}[H]
    \centering
    \includegraphics[width=1\linewidth]{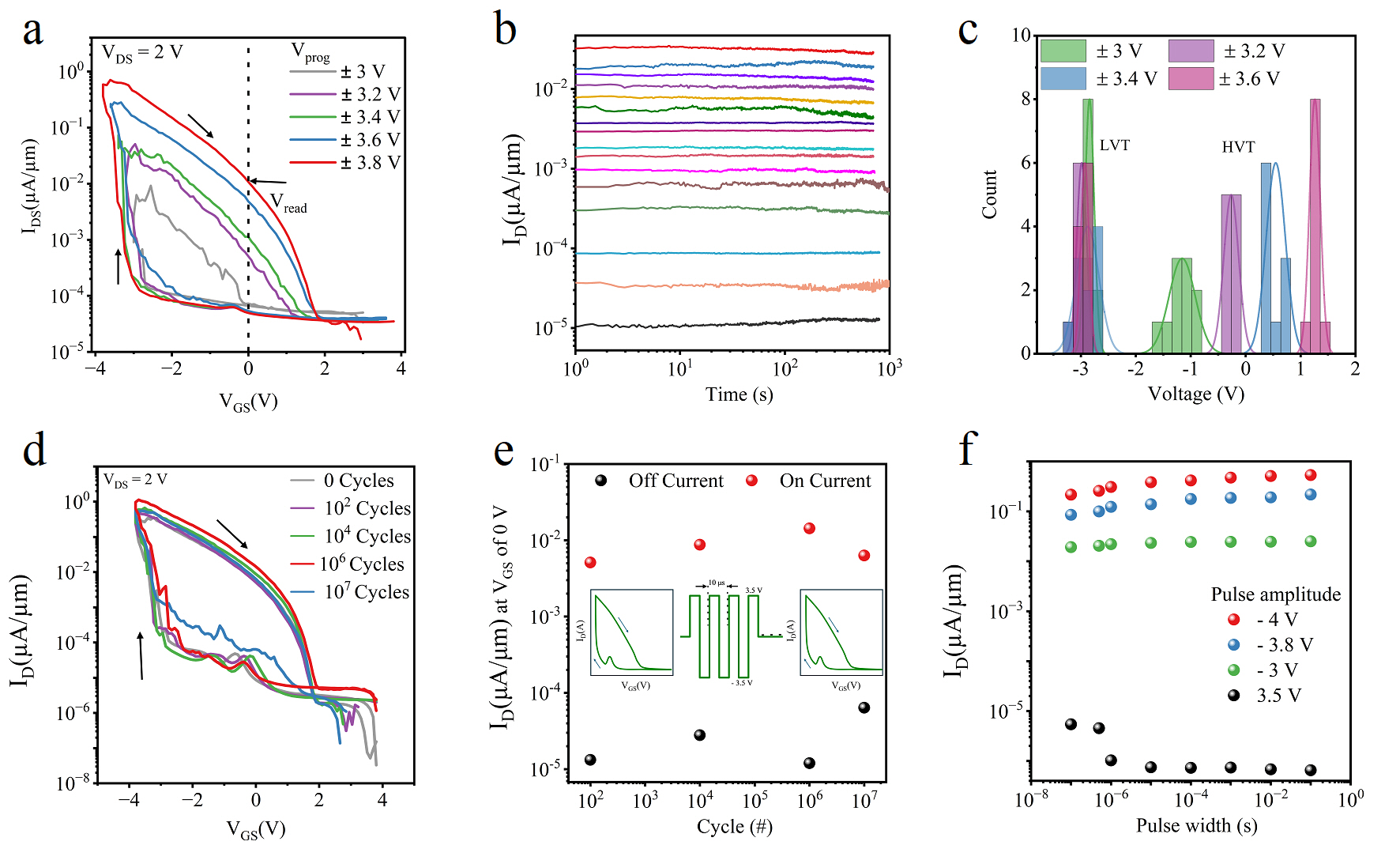}
    \caption{Switching speed, endurance, and retention behavior of the MoS$_2$ p-FeFET. (a) Transfer characteristics after the wake-up process for $V_{\mathrm{GS}}$ sweep ranges of $\pm 3.0$, $\pm 3.2$, $\pm 3.4$, $\pm 3.6$, and $\pm 3.8$~V.(b) Retention characteristics of different programmed states measured at $V_{\mathrm{GS}}=0$~V and $V_{\mathrm{DS}}=0.5$~V over 300~s. (c) Cycle-to-cycle variation of LVT and HVT extracted from 10 repeated transfer characteristic measurements at each $V_{\mathrm{GS}}$ sweep range of $\pm 3.0$, $\pm 3.2$, $\pm 3.4$, and $\pm 3.6$~V. (d) Evolution of the transfer characteristics after applying bipolar $\pm3.5$~V gate-bias pulses with a period of 10~$\mu$s. The transfer characteristics were monitored after every two decades of stress cycles. (e) Extracted $I_{\mathrm{ON}}$ and $I_{\mathrm{OFF}}$ at $V_{\mathrm{GS}}=0$~V after every two decades of stress cycles; the inset shows the endurance measurement protocol. (f) Programming speed measured using negative gate pulses with amplitudes of $-3$, $-3.8$, and $-4$~V and pulse widths ranging from 500~ns to 100~ms. The erase speed was measured using a 3.5~V gate pulse with pulse widths ranging from 500~ns to 100~ms.}
    \label{fig:7}
\end{figure}

Furthermore, the memory performance of the MoS$_2$ p-FeFET was evaluated through zero-gate bias retention, cycle-to-cycle (C2C) variation, and endurance measurements. Figure~\ref{fig:7}a shows the transfer characteristics after wake-up over different $V_{\mathrm{GS}}$ sweep ranges. Figure~\ref{fig:7}b presents the retention characteristics of the programmed states. For each measurement, the device was first set to the OFF state by applying a positive $V_{\mathrm{GS}}$ pulse of 3.8~V for 1~ms. A negative programming pulse was then applied to set the device to the desired state, followed by drain-current readout at $V_{\mathrm{GS}}=0$~V and $V_{\mathrm{DS}}=0.5$~V over time. 

By varying the amplitude and width of the negative programming pulses, 16 distinct programmed states corresponding to 4-bit memory were obtained, spanning more than three orders of magnitude in drain current with clear separation between states. All states remained distinguishable over the measured 700~s without signs of degradations, demonstrating the nonvolatile memory behavior of the p-FeFET. The C2C distributions of the LVT and HVT states at different $V_{\mathrm{GS}}$ sweep ranges are presented in Fig.~\ref{fig:7}c. The LVT distributions overlap for the different sweep ranges, whereas the HVT distributions exhibit distinct mean values with narrow spreads. These results demonstrate the ability to obtain multiple distinguishable threshold-voltage states in the device. Figure~\ref{fig:7}d display the transfer characteristics after repeated gate bias stress. The pulsing protocol used for the bias stress is shown in the inset of Figure~\ref{fig:7}e, where bipolar $\pm3.5$~V square pulses with a period of 10~$\mu$s were applied for up to $10^{7}$ cycles. The transfer characteristics were monitored after every two decades of pulse cycles. The transfer curve remains largely unchanged up to $10^{6}$ cycles. After $10^{7}$ cycles, a slight increase in the OFF-state current and a reduction in the ON-state current are observed (Figure \ref{fig:7}e) These results demonstrate better stability of the device under repeated gate bias stress compared to Si-FeFETs. The degradation after $10^{7}$ switching cycles is associated with fatigue induced at the HZO stack due to prolonged bias stressing, as verified by the gate stack P-V loop and I-V endurance (Supplementary information, Figure S6). This effect has been studied in our previous work thoroughly showing that ferroelectric fatigue originates from cycling of oxygen vacancy movement under prolonged bias stressing in HZO thin films and can be improved with proper gate oxide engineering.\cite{li2025understanding} Further improvement of the gate stack with higher endurance is currently underway.

Next the programming and erasing speeds were examined by varying the gate-pulse width. For the programming-speed measurement, the device was first initialized to the OFF state using a positive \(V_{\mathrm{GS}}\) pulse of 3.5 V for 1 ms. A negative programming pulse was then applied, with the pulse width varied from 100 ns to 100 ms. After each programming pulse, the drain current was read at \(V_{\mathrm{GS}}=0\) V and \(V_{\mathrm{DS}}=0.5\) V. The device was reset to the OFF state before applying each subsequent programming pulse. This measurement was repeated for programming-pulse amplitudes of \(-3\), \(-3.8\), and \(-4\) V. The detailed pulsing protocol is provided in the Supplementary Information Figure S7. Figure~\ref{fig:7}f shows the drain current as a function of pulse width for different programming amplitudes. At a programming amplitude of \(-3\) V, the drain current increases by more than three orders of magnitude for a pulse width of 100 ns. With increasing pulse width, the drain current continues to rise up to approximately 100 \(\mu\)s, with a change of less than one order of magnitude, before reaching saturation. A similar trend is observed at programming amplitudes of \(-3.8\) and \(-4\) V, with higher programming amplitudes yielding higher drain currents. These results demonstrate that our MoS$_2$ p-FeFET can be programmed using pulse widths as short as 100 ns. The corresponding I\textsubscript{D}-t measurements for each programming pulse are provided in Supplementary Figures S8–S10. The erase speed was evaluated by first setting the device to the ON state using a programming pulse of \(-3.8\) V for 100 \(\mu\)s. A positive erase pulse of 3.5 V was then applied with pulse widths ranging from 100 ns to 100 ms. Before each erase pulse, the device was returned to the ON state using the same programming conditions. The drain current was measured after each erase pulse at \(V_{\mathrm{GS}}=0\) V and \(V_{\mathrm{DS}}=0.5\) V. As shown in Fig.~\ref{fig:7}f, the drain current remains above the fully erased level for an erase-pulse width of 100 ns and decreases with increasing pulse width, reaching the fully erased level at 10~$\mu$s. This indicates that erasing requires longer pulses than programming. The pulsing scheme used for the erase-speed measurements and the corresponding I\textsubscript{D}-t characteristics are presented in Supplementary Figures S11 and S12. The slower erase response is consistent with the transfer characteristics, where the drain current rises sharply as the gate voltage is swept from positive to negative but falls more gradually during the reverse sweep. Lack of enough electrons in the MoS$_2$ channel to erase the hole-rich state is the reason for the slower erase process that needs further work. A comparison of MoS$_2$-based FeFETs employing different ferroelectric gate dielectrics is presented in Table~\ref{tab:comparison}. Compared with previously reported devices, our MoS$_2$/HZO FeFETs offer a combination of low-voltage operation, a large memory window, high endurance and multibit memory functionality. 
  
\begin{table*}[!t]
\centering
\caption{Comparison of MoS$_2$-based FeFETs employing different ferroelectric gate dielectrics.}
\label{tab:comparison}

\small
\setlength{\tabcolsep}{3pt}
\renewcommand{\arraystretch}{2.1}

\begin{tabular}{|
>{\centering\arraybackslash}p{3.1cm}|
>{\centering\arraybackslash}p{1.0cm}|
>{\centering\arraybackslash}p{1.6cm}|
>{\centering\arraybackslash}p{1.4cm}|
>{\centering\arraybackslash}p{1.7cm}|
>{\centering\arraybackslash}p{1.7cm}|
>{\centering\arraybackslash}p{1.5cm}|
>{\centering\arraybackslash}p{1.5cm}|
>{\centering\arraybackslash}p{1.7cm}|}
\hline

\textbf{Device stack\textsuperscript{\large *} [Ref]} &
\textbf{MoS$_2$} &
\textbf{FE} &
\makecell{\textbf{FE thi-}\\\textbf{ckness}} &
\makecell{\textbf{$V_{\mathrm{P}}$/}\\\textbf{MW}} &
\makecell{\textbf{$I_{\mathrm{ON}}$/}\\\textbf{$I_{\mathrm{OFF}}$}} &
\makecell{\textbf{Reten-}\\\textbf{tion}} &
\makecell{\textbf{Endur-}\\\textbf{ance}} &
\makecell{\textbf{Memory}\\\textbf{states}}
\\
\hline

HZO/MoS$_2$ FeFET \textsuperscript{\large *}~\cite{zhang2020low}
&
Few-layer
&
HZO
&
6 nm
&
$\pm3$ V/0.1 V
&
$>10^{7}$
&
$10^{4}$ s
&
$10^{3}$
&
2
\\
\hline

Al:HfO$_2$/MoS$_2$ FeFET \cite{10.1063/1.4991877}
&
1 L
&
Al-doped HfO$_2$
&
16 nm
&
$\pm10$ V/0.125 V
&
--
&
--
&
$2\times10^{4}$
&
2
\\
\hline

P(VDF-TrFE)/MoS$_2$\textsuperscript{\large *} \cite{https://doi.org/10.1002/aisy.202300391} 
&
1 L
&
P(VDF-TrFE)
&
30 nm
&
$\pm 5$ V/3 V
&
$>10^{6}$
&
$>10^{3}$ s
&
--
&
$16$ 
\\
\hline

PZT/MoS$_2$ FeFET \cite{Sun_2017}
&
4 L
&
PZT
&
260 nm
&
$\pm 10$ V/2.5 V
&
--
&
--
&
--
&
---
\\
\hline

P(VDF-TrFE)/MoS$_2$ \cite{aspiotis2023large}
&
1 L
&
P(VDF-TrFE)
&
30 nm
&
$\pm5$ V/3 V
&
$10^{7}$
&
$10^{3}$ s
&
--
&
2
\\
\hline

BLFMO/MoS$_2$ FeFET~\cite{zhang2022tunable}
&
$\sim$19 L
&
BLFMO
&
$\sim$200 nm
&
$\pm25$ V/$>25$ V
&
$\sim10^{5}$
&
$10^{3}$ s
&
$10^{3}$
&
4
\\
\hline

HZO/MoS$_2$ FeFET \textsuperscript{\large *} ~\cite{10413873}
&
$\sim$1 L
&
HZO
&
$\sim$2.5 nm
&
$\pm1$ V/$0.23$ V
&
$10^{3}$
&
$\geq10^{3}$ s
&
$10^{12}$
&
2
\\

\hline

HZO/MoS$_2$ FeFET \textsuperscript{\large *} ~\cite{10.1021/acsnano.6c08339}
&
$\sim$1 L
&
HZO
&
$\sim$11 nm
&
$\pm3.5$ V/$3.6$ V
&
$10^{5}$
&
$10^{4}$ s
&
$10^{5}$
&
2
\\

\hline

AlScN/MoS$_2$ FeFET \textsuperscript{\large *} ~\cite{kim2023scalable}
&
$\sim$1 L
&
AlScN
&
$\sim$45 nm
&
$\pm10$ V/$7.8$ V
&
$10^{7}$
&
$\geq10^{5}$ s
&
$10^{4}$
&
16
\\

\hline
 HZO/MoS$_2$ FeFET \textsuperscript{\large *} 
 
 \textcolor{blue}{(this work)}
&
 $\sim$3 L
&
 HZO
&
 $\sim$20 nm
&
$\pm3.8$ V / $4.4$ V
&
 $\sim10^{5}$
&
 $\geq10^{3}$ s
&
 $10^{7}$
&
 16
\\

\hline

\end{tabular}
\vspace{2pt}
\begin{flushleft}
\footnotesize
\textsuperscript{\large *} CMOS BEOL-compatible devices.
\end{flushleft}

\end{table*}

This benchmarking highlights the potential of our integration strategy for next-generation ferroelectric memory devices, combining competitive memory performance with CMOS BEOL-compatible processing. 
\clearpage

\section{Conclusion}
In conclusion, we demonstrated a promising BEOL compatible p-type MoS\textsubscript{2} FeFET with large memory window of 4.4 V at an operating voltage of $\pm$3.8 V and 16 programmable memory states with non-volatile retention enabled by wafer-scale ALD-grown MoS\textsubscript{2} as the channel material and HZO as the ferroelectric gate oxide. This approach where polarization-induced modulation of the electrostatic potential at the MoS\textsubscript{2}/contact interface plays an important role in controlling hole transport, achieved an excellent device to device performance uniformity and on/off current ratio exceeding $10^4$. The devices exhibit endurance beyond $10^7$ switching cycles and programming times as short as 100 ns. This work shows 2D FeFETs as a promising technology toward scalable, BEOL-compatible p-type  multibit non-volatile memory contributing towards the advancement of two-dimensional CMOS technology for next-generation electronics.

\section{Experimental Section}
\label{experimental_section}
\subsection{Growth of  monolayer MoS$_2$} 
The substrates used for the growth of the MoS$_2$ films are 6-inch p-type silicon wafers. First, a 300 nm wet-thermal SiO$_2$ layer is grown at 1000~$^\circ$C in a tube furnace. The wafers are then treated in a UV/O$_3$ reactor for 10 min to improve the chemical termination of the oxide surface. MoO$_3$ is subsequently deposited by thermal atomic layer deposition (ALD) using a Veeco Savannah S200 system, with bis(tert-butylimido)bis(dimethylamido)molybdenum as the molybdenum precursor and ozone as the oxidant. Fifteen ALD cycles at 250~$^\circ$C produce an approximately 1.4 nm thick MoO$_3$ film with high uniformity across the entire 6-inch wafer. This initial deposition step provides precise control over the number of MoS$_2$ layers formed during subsequent processing. Following deposition, the MoO$_3$ layer is converted to MoS$_2$ by annealing in a tube furnace. To accommodate the furnace tube dimensions, the wafer is diced into 40 mm $\times$ 150 mm chips. The chips are placed on a crucible and loaded into the furnace for sulphurisation in an H$_2$S/Ar atmosphere. The process comprises two stages: conversion of MoO$_3$ to MoS$_2$ at 300~$^\circ$C, followed by annealing at 970~$^\circ$C to crystallise the film. After conversion and crystallisation, the furnace is allowed to cool naturally to room temperature. This process produces a uniform MoS$_2$ film directly on SiO$_2$, which can be used without transfer to fabricate high-quality rigid electronic devices.
  
\subsection{Structural Characterization}  

\subsubsection{X-ray diffraction (XRD)}
X-ray diffraction (XRD) measurements of similar Hf$_{0.5}$Zr$_{0.5}$O$_2$ stacks showed that annealing at 350~$^\circ$C slightly increases the fraction of the non-polar monoclinic phase, while a significant fraction of the ferroelectric orthorhombic phase remains (Supplementary Fig.~S2).\cite{11481754}  

\subsubsection{Raman Spectroscopy} 
Raman spectroscopy was used to assess the quality of the 15-cycle ALD MoS$_2$ film on 300 nm SiO$_2$/Si substrates using a Renishaw Ramascope with a 532 nm excitation laser. The spectrum exhibited two characteristic MoS$_2$ peaks at 382.128 cm$^{-1}$ and 405.806 cm$^{-1}$, corresponding to the E$^1_{2g}$ in-plane and A$_{1g}$ out-of-plane phonon modes, respectively. The separation between these peaks ($\Delta$) provides an estimate of the number of MoS$_2$ layers. The measured separation of 23.678 cm$^{-1}$ indicated a film thickness corresponding to 2--3 MoS$_2$ layers.

\subsubsection{X-ray Photoelectron spectroscopy(XPS)} 
X-ray photoelectron spectroscopy (XPS) was performed using a non-monochromatized RS 40B1 Al K$\alpha$ X-ray source (Prevac) and an Argus electron spectrometer (Omicron Nanotechnology), with a sampling area of $\varnothing 1.93$ mm$^2$. Spectra were acquired at normal emission to increase the sampling depth. All measurements were conducted under ultra-high vacuum conditions with a base pressure of approximately $1 \times 10^{-10}$ Torr. The binding energy scale was calibrated to 284.8~eV using the C~1s peak.
\subsection{Device fabrication and electrical Measurement} 
Electrical characterization of the samples was carried out by applying voltage between the bottom Ti/Pt and the top gold electrode for the MFIM capacitor. For the FeFET, drain-to-source voltage ($V_{\mathrm{DS}}$) was applied between the drain (D) and source (S) contacts, while gate voltage ($V_{\mathrm{G}}$) was applied at the Pt bottom electrode. For ferroelectric polarization-voltage ($P$--$V$) hysteresis, endurance and retention measurements, an AixACCT 2000E ferroelectric material tester was used. Static current-voltage ($I_{\mathrm{D}}$--$V_{\mathrm{D}}$) and $I_{\mathrm{D}}$--$V_{\mathrm{G}}$ characteristics were measured using a Keysight 1500A semiconductor parameter analyzer. Ferroelectric hysteresis was measured on both MFIM and MFIS structures (between the gate and source contacts) using dynamic hysteresis measurement (DHM) techniques with a triangular voltage sweep at 1~kHz. For endurance measurements, rectangular voltage pulses of 4~V amplitude and 100~kHz frequency were applied, followed by one $I_{\mathrm{D}}$--$V_{\mathrm{G}}$ measurement per decade to monitor the ON/OFF ratio and memory window of the samples over prolonged bias stressing. All measurements were carried out at room temperature and under ambient conditions.

For retention measurements, a programming pulse of different magnitude or an erase pulse was first applied at the gate, and the drain current was monitored over time at $V_{\mathrm{G}} = 0$~V. Switching-speed studies were performed by applying a single pulse with different pulse widths (from the millisecond to the 100~nanosecond range) and different pulse magnitudes. First, the device was set to the OFF state, and the current response to a single pulse was recorded to determine how quickly the ON current level could be reached and thereby evaluate the programming speed. For erase-speed measurements, a similar protocol was followed by initially setting the device to the ON state and then applying a single pulse to estimate how quickly the device could reach the OFF state.

\section{Acknowledgment}
The authors M.R., F.A.S., and S.M. gratefully acknowledge the financial support provided by the Research Council of Finland (RCF) through the SA/Profi 8 (Grant No. 312136522512). F.A.S. also acknowledge the financial support provided by the Finnish Ministry of Education and Culture through the PREIN/I-DEEP Doctoral Pilot (VN/3137/2024-OKM-4) and the Flagship for Photonics Research and Innovation (PREIN) decision number Research Council of Finland/2024/368650 Tampere University. We acknowledge OtaNano Micronova cleanroom, SI Cleanroom of Tampere University for fabrication support and Laboratory of Future Electronics and ChipIn infrastructure of Tampere University for characterization of the samples.  

\bibliography{ref}
\end{document}